\documentclass[
 reprint,
 superscriptaddress,
showpacs,preprintnumbers,
 amsmath,amssymb,
 letter,
 aps,
 prl
]{revtex4-1}

\usepackage{graphicx}
\usepackage{subfigure}
\usepackage{float}
\usepackage{color}
\usepackage{dcolumn}
\usepackage{bm}
\usepackage{dblfloatfix}
\usepackage{fixltx2e}
\usepackage{tikz}
\usetikzlibrary{arrows}

\begin{document}


\title{Consolidating Multiple FemtoSecond Lasers in Coupled Curved Plasma Capillaries}

\author{A Zigler}
\author{M Botton}
\email{bdmoti@mail.huji.ac.il}
\affiliation{
 Hebrew University of Jerusalem, Jerusalem 91904, Israel
}
\author{F Filippi}
\affiliation{%
 Laboratori Nazionali di Frascati, INFN, Via E. Fermi, Frascati, Italia
}
\author{Y Ferber}
\author{G. Johansson}
\author{O Pollack}
\affiliation{
	Hebrew University of Jerusalem, Jerusalem 91904, Israel
}
\author{M.P. Anania} 
\author{F. Bisesto} 
\author{R. Pompili}  
\author{M. Ferrario}
\affiliation{%
	Laboratori Nazionali di Frascati, INFN, Via E. Fermi, Frascati, Italia
}
\author{E Dekel}
\affiliation{
	Hebrew University of Jerusalem, Jerusalem 91904, Israel
}

\date{\today}

\begin{abstract}
Consolidating multiple high-energy femtosecond scale lasers is expected to enable implementation of cutting edge research areas varying from wakefield particle accelerators to ultra-high intensity laser pulses for basic fresearch. The ability to guide while augmenting a short-pulse laser is crucial in future laser based TeV particle accelerators where the laser energy depletion is the major setback. We propose, analyze and experimentally demonstrate consolidating multiple femtosecond pulse lasers in coupled curved capillaries. We demonstrate a proof of principle scheme of coupled curved capillaries where two femtosecond laser pulses are combined. We found that the details of the coupling region and injection scheme are crucial to the pulse consolidations. Furthermore, our simulations show that high-intensity short pulse laser can be guided in a small curvature radius capillary. Incorporating these finding in a curved capillary laser coupler will be a significant step towards realization of meters long TeV laser based particle accelerators.


\end{abstract}

\maketitle

High intesity short pulse lasers have opened new horizons in many research aspects of light-matter interaction\cite{ledingham2010review}. Laser systems of up to several PW level were demonstrated and utilized in experiments. Research areas ranging from astrophysics-in-lab to TeV laser based particle accelerators are the main beneficiaries of the novel high intensity laser physics \cite{BORGHESI2014review,Bulanov2015review,corde2013review,daido2012review,esarey2009review}. Nevertheless, generation of PW scale pusles from a single source turned out to be exceedingly complicated task that requires both specifically engineered expensive components as well as expertise, limiting the number of operating systems \cite{danson_hillier_hopps_neely_2015}. Medium scale energy lasers (100TW) have become the working tool in many university roomsize research labs. It is therefore expected that combining pulses from such systems can fill the gap and provide a much wider accessability to PW laser pulses. On a different aspect, guided laser pulse in a plasma channel that is created in a capillary was shown to accelerate electrons up to GeV energy levels \cite{leemans2006gev,Kim2013multigev,leemans2014multigev} utilizing laser wakefield acceleration scheme (LWFA). Comparing that to the kilometer-length conventional accelerators it is clear why the LWFA is considered to be a most promissing option for future high energy particle accelerators. The three main limitations of LWFA are laser beam diffraction, electron dephasing, and laser energy depletion \cite{esarey2009review}.  It has been demonstrated in the past by many groups that a hollow radial plasma density profile enables to extend the length along which the beam remains sufficiently focused to up to tens of centimeters \cite{ehrlich1996, zigler1996optical,levin2006long}. Electron dephasing occurs when the electron bunch begins to slip ahead of the range in the wakefield where they are accelerated. We have also previously shown that tapering the plasma density along the direction of propagation can postpone electron dephasing \cite{kaganovich1999variable,kaganovich2001velocity}. Energy depletion of the laser beam is commonly envisaged as being dealt with by concatenating several acceleration stages. Multi-staging is expected to avoid most of these problems, allowing to locally inject a different, undepleted laser pulse into the plasma channel which can be independently delayed to rephase the driver with the accelerating electrons. Multi-stage scheme has been recently proved \cite{Steinke2016} using plasma mirrors to inject the laser into preformed plasma channels and plasma lens to capture the laser generated electrons before sending them into the subsequent acceleration stages. This scheme sets a significant distances between the stages and has the drawback of a low coupling efficiency far below the suitable for future accelerators. A possibility of adding a second laser through a curved capillary was suggested several years ago \cite{zigler_curvcap2016} and recently simulated \cite{Lou2018curvedcap}.

\begin{figure}[H]
	\begin{center}
		\includegraphics[scale = 0.36, trim= 0 0 0 0 ]{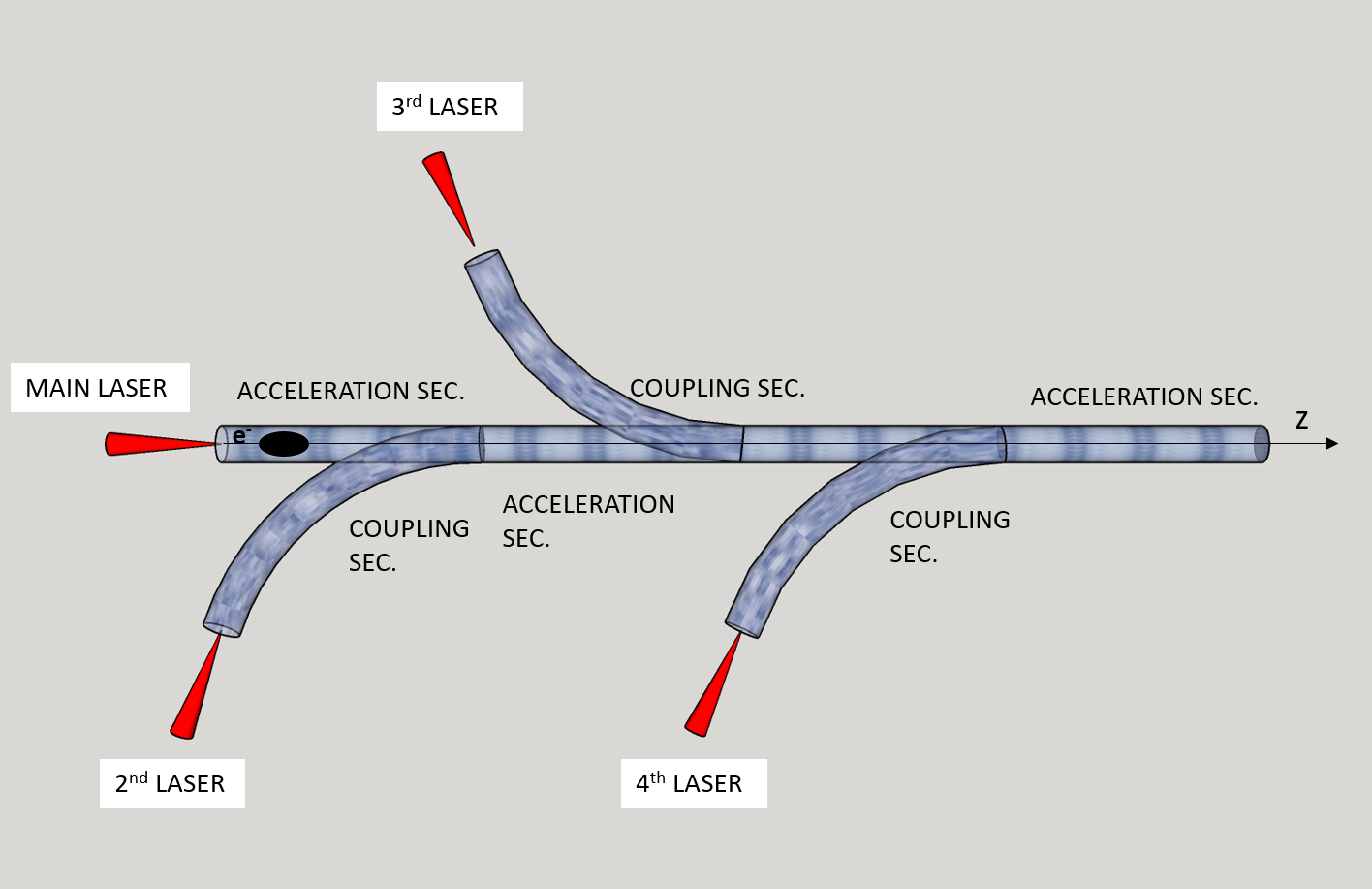}		
        \begin{picture}(0,0)
            \put(0,105){\includegraphics[height=2.2cm]{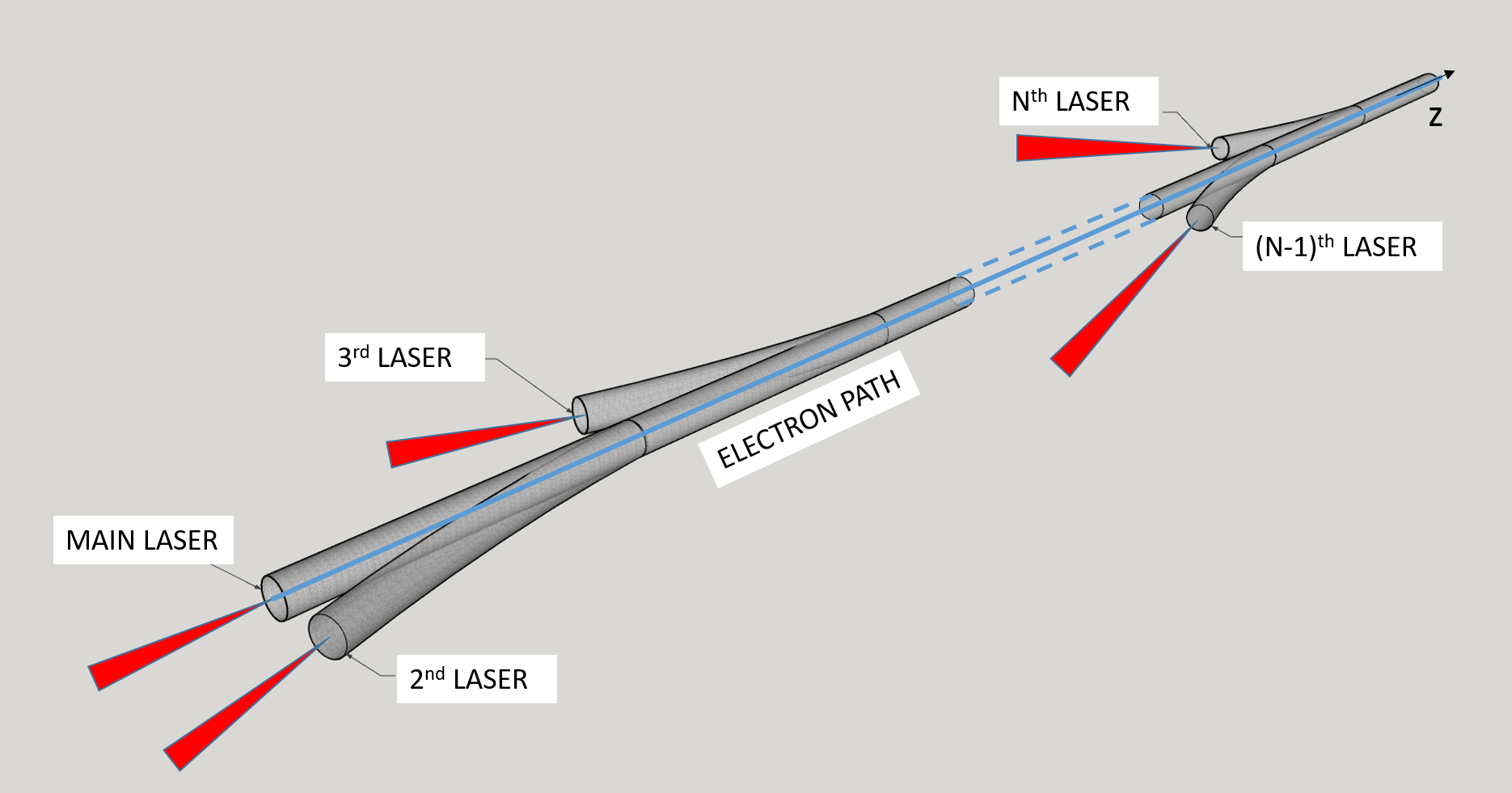}}
        \end{picture}
		\caption{Proposed multistage acceleration scheme. The first main laser is initiating the acceleration of the elctron beam. A Fishbone shaped curved channels are added dpwn the acceleratoin line enabling further acceleration of the electrons. A 3D concept of the fishbone accelerator is shown in the inset. }
		\label{StagesSchem}
	\end{center}
\end{figure}

\clearpage

\onecolumngrid
\begin{figure*}[t]
	\begin{center}
		\includegraphics[scale = 0.40, trim= 0 0 0 0 ]{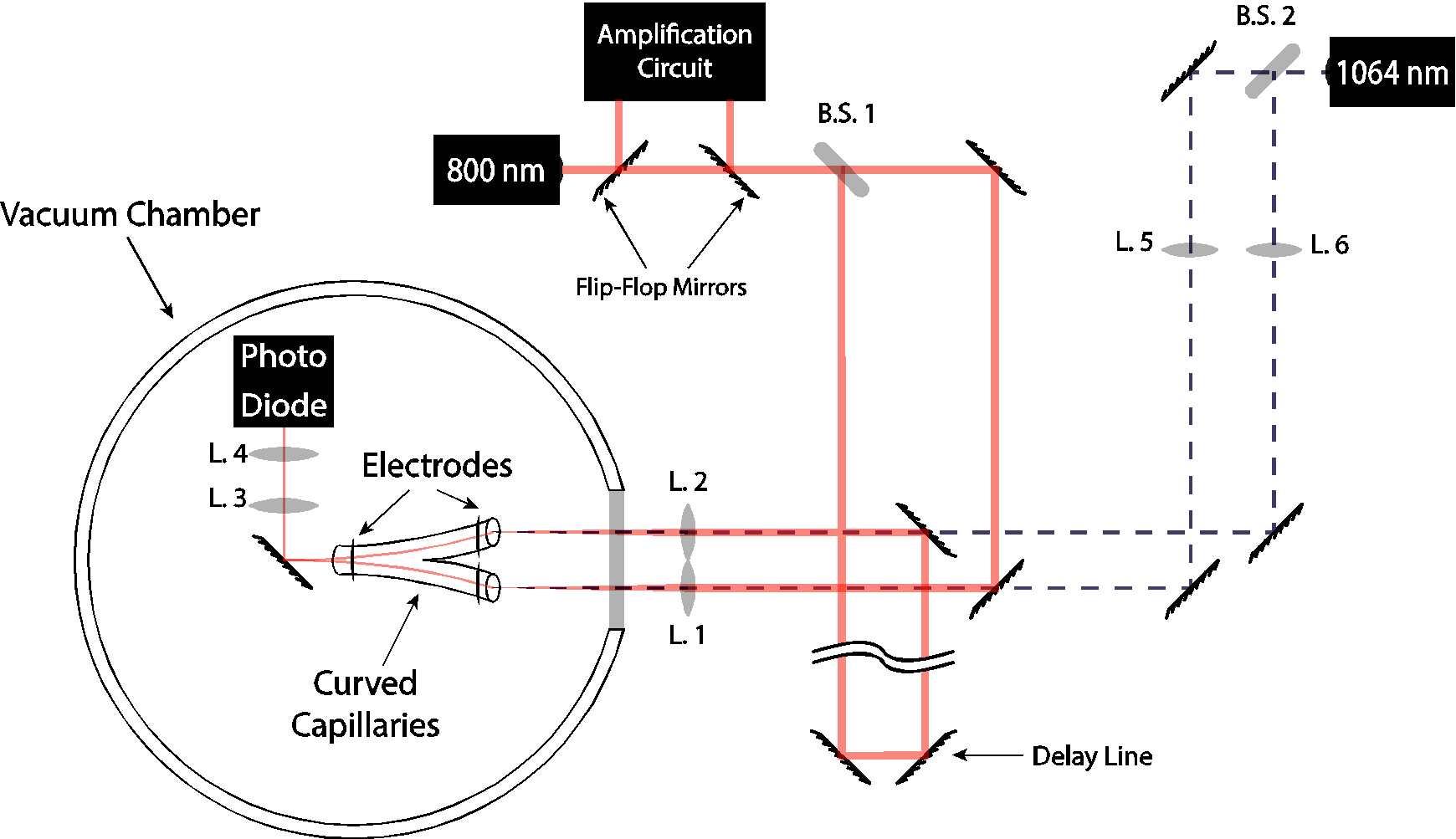}
		\caption{Experimental setup.The Y shaped coupled capillaries junction is in the vacuum chamber. Main laser line is shown in red. The triggering laser line is in broken line. The coupled capillaries }
		\label{ExpSetup}
	\end{center}
\end{figure*}
\twocolumngrid
In this letter, we demonstrate a novel concept of concatenating consecutive acceleration stages. The proposed concept is based on a "Fish-Bone" like structure (see Fig. \ref{StagesSchem} for a possible realization). The main laser accelerates an electron bunch by the LWFA scheme, till the laser energy depletion.  Then, curved channels allow to add new properly timed laser pulses on the acceleration axis driving the wakes to further accelerate the bunch. With this scheme the overal length of a TeV accelerator can be reduced by at least an order of magnitude   as the accelerator units are closely packed. Nevertheless, there are two major issues that must be addressed in order to implement this innovative scheme. 

First, the coupling region of the two capillaries must be carefully designed so that the laser pulses will be properly joined while the electron bunch will not be broken. Secondly, the injecting cappilary should be curved to the smallest possible radius in order to shorten the injection area. Addressing the first and formost issue, we have conducted a proof of principle experiment by coupling two laser beams. The scheme implemented in this proof of principle is a "Y" shaped plasma channel coupling region (see Fig. \ref{ExpSetup}). Each branch comprises a curved capillary filled with a plasma of a radial density profile. The capillaries inner radii is 250$\mu$m, and the radius of curvature is 100 cm.  The capillaries Y-junction was manufactured by a 3D-printing system with spatial resolution of 12$\mu$m \cite{FilippiTBP}. The prining material was chosen to satisfy the experimental requirements of the system. In particular, the material was desinged to be hard in order to reduce over ablation and transparrent to facilitate light based diagnostics of the plasma channel. Accordingly, a $CH_2$ based material were used.

The generation of the plasma channel is based on the laser trigger ablative technique \cite{palchan2007electron}. An alternative approach of gas driven capillaries is obviuosly possible as well.
The electronic part of the discharge mechanism consists of a high-voltage source, a capacitor for storing the energy to be discharged and a resistor for limiting the charging current. The anode and cathode are two electrodes placed at either end of the capillary. The triggering is obtained by a 1064 nm laser of 10 ns pulses with 50 mJ per pulse. The triggering beam is splitted by a beam splitter to timely trigger each branch (simultaneously or with a predetermined time delay). Following the trigger, an avalanch ablative process is started and due to the discharge current a plasma is formed in the capillaries. For an electric discharge of $\sim$15 kV along the 5cm of the capillary, a plasma with an average density of $5 \cdot 10^{17}$ cm$^{-3}$ is formed with an approximately hollow RDP which is suitable for the laser guiding. This isn't far from the optimal plasma density for LWFA of $2 \cdot 10^{18}$ cm$^{-3}$ which can be easily reached by reducing the capillary radius or increasing the current furnished by the circuit \cite{kaganovich1999variable}.
The main beam to be guided, (red line in Fig. \ref{ExpSetup}), originates from a 800 nm laser, which produces 5 nJ, 12.5 fs pulses, wihch can be amplified to a pulse of 5 mJ, 30 fs. This beam is split as well (B.S. 1) for the two capillaries, with one of the branches going through a delay line, after which they each reach a lens (L.1,2) and are focused into the capillaries. At the exit of the capillaries, an imaging system (L. 3,4) enables the intensity measurment of the transmitted light.  The lifetime of the plasma channel (with its proper density parameters) is approximately 100ns. Accordingly the timing of the channel generation and the main laser pulse injection is a crucial parameter. Utilizing the laser triggering technique with a time jitter of about 20ns is essential to meet these strict timing requirement

\begin{figure}[h]
	\begin{center}
		\includegraphics[scale = 0.37, trim= 0 0 0 0 ]{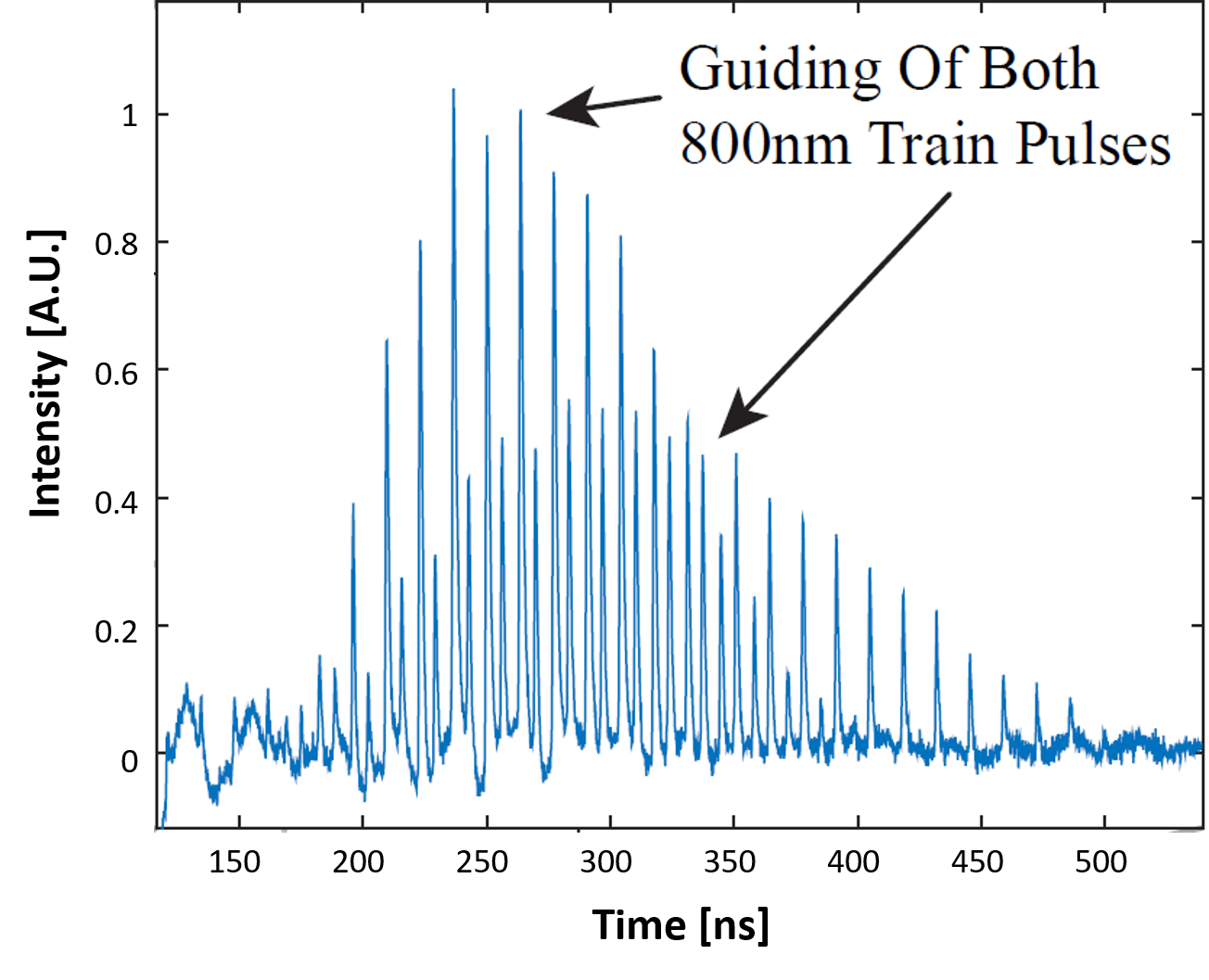}
		\caption{Experimental demonstration of a simultaneous guiding of two low power laser pulses. Each pulse is injected toa different branch of the coupled capillaries. The combined pulse is measured at the exit branch.}
		\label{GuidingWindow}
	\end{center}
\end{figure}

A clear experimental demonstration of consolidation of two laser pulses in the curved capillary Y junction is shown in Fig. \ref{GuidingWindow}. Two 800nm pulse trains are injected into the two separate entry points of the capillaries. Each pulse in the train contains 5nJ and is of 12.5fs duration. The pulse repetition rate is 76GHz and a 5ns time delay is introduced between the two branchews in order to distinguish one from the other. The output enerrgy is measured at the oputput of the combined capillary and is shown in the figure. It is clear that both pulses propagates through the junction with a considerable efficiency. Furthermore, the time duration of the guiding window which starts about 250ns after the trigger pulse, is clearly shown in Fig. \ref{GuidingWindow}. 

\begin{figure}[h]
	\begin{center}
		\includegraphics[scale = 0.42, trim= 0 0 0 0 ]{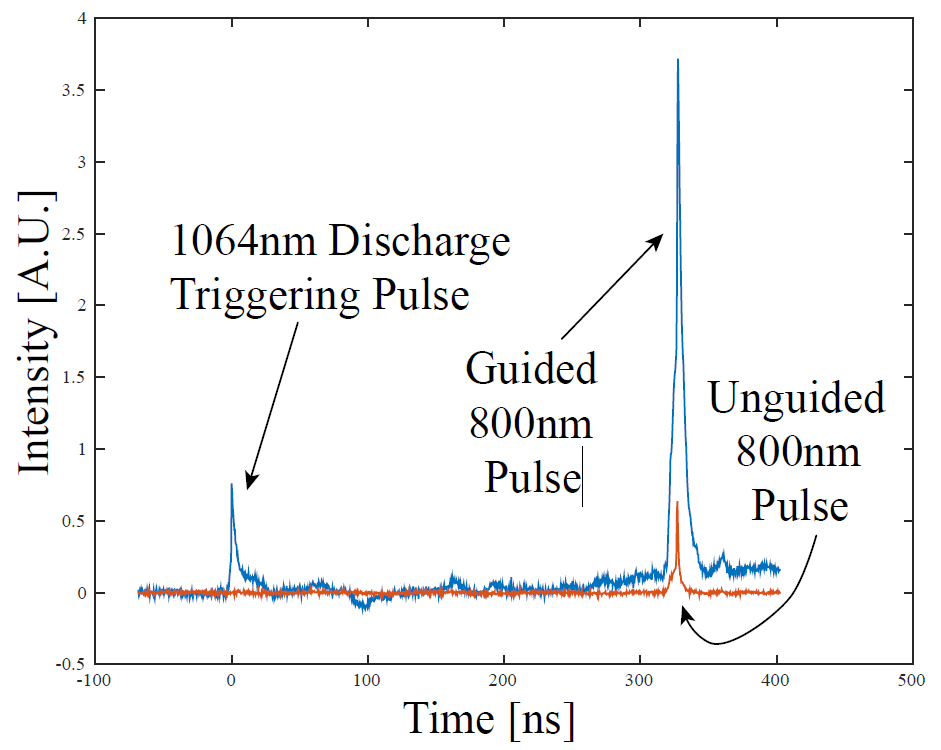}
		\caption{Guiding of high peak power laser pulse in a curved capillary.The laser is injected to one branch of the coupled capillaries structure where both branched are triggered to create plasma channel. The guided puulse is measured at the exit branch.}
		\label{GuidingRes}
	\end{center}
\end{figure}

A demonstration of the guiding effect of a high power beam is given in Figure \ref{GuidingRes}. The exit magnitude of the guided high intensity (0.1TW) pulse is 6 times stronger with the guiding effect compared to a non guiding conditions where the main pulse inefficiently drifts through the curved capillary by moltiple internal reflections.

\begin{figure}[H]
	\begin{center}
		\includegraphics[scale = 0.3, trim= 0 0 0 0 ]{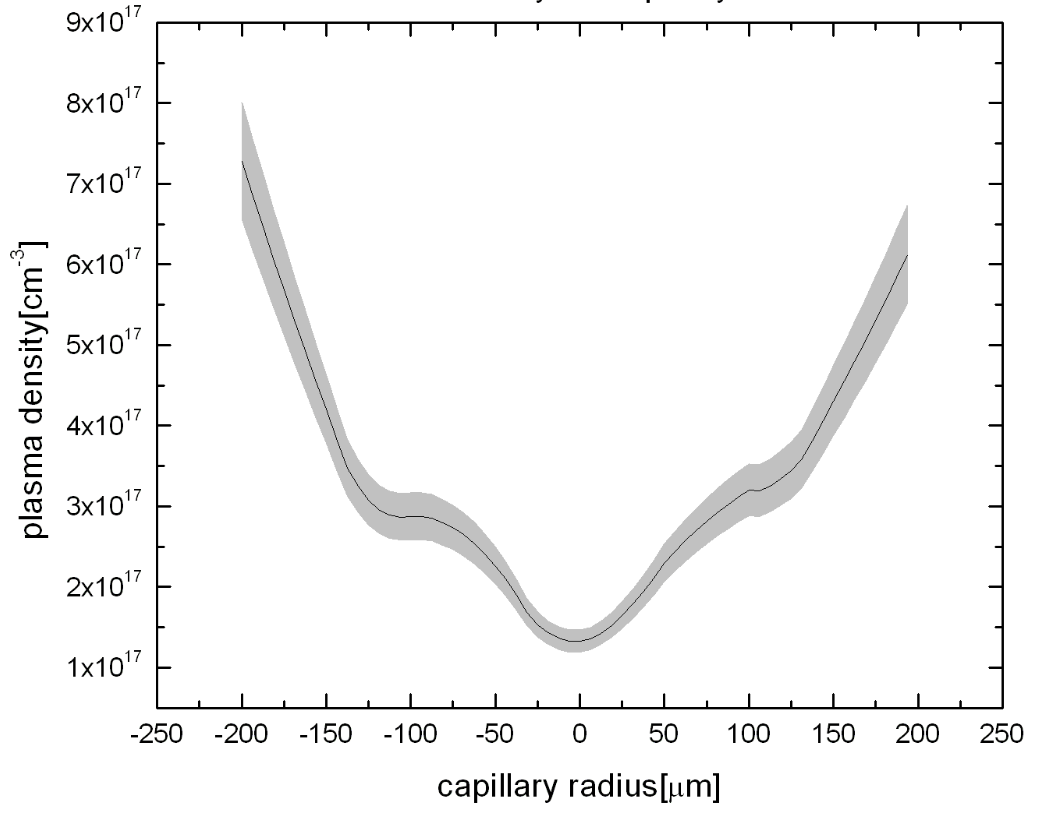}
        \begin{picture}(0,0)
           \put(-160,100){\includegraphics[height=2.5cm]{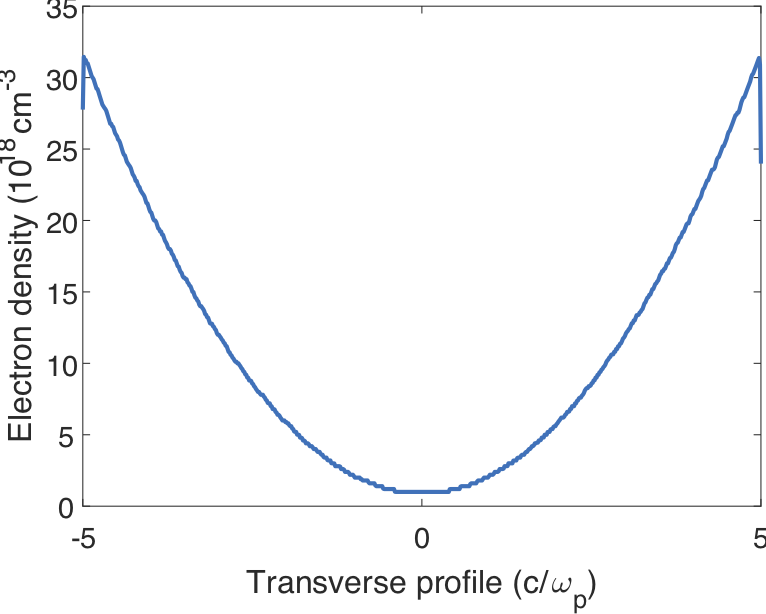}}
        \end{picture}
		\caption{Plasma radial density profile in the capillary measured at the entrance of the capillary. The simulated progfile is shown in the inset.}
		\label{TransverseProfile}
	\end{center}
\end{figure}

Having experimentally established the consolidation of two pulses in a curved capillary Y junction and the transmission of high intensity short pulse in a curved capillary  
we proceed to tackle the second issue, namely significantly reducing the curvature radius of each branch. Here we propose a numerical simulation study of possible parameters to reach the desired goal. The simulations are based on the TURBOWAVE code \cite{gordon_tw1,gordon_tw2}. Fig. \ref{TransverseProfile} presents the measured plasma density in a capillary discharge. The inset shows the density profile used in the simulations. 

\clearpage
\onecolumngrid
\begin{figure*}
	\minipage{0.32\textwidth}%
	\includegraphics[width=\linewidth]{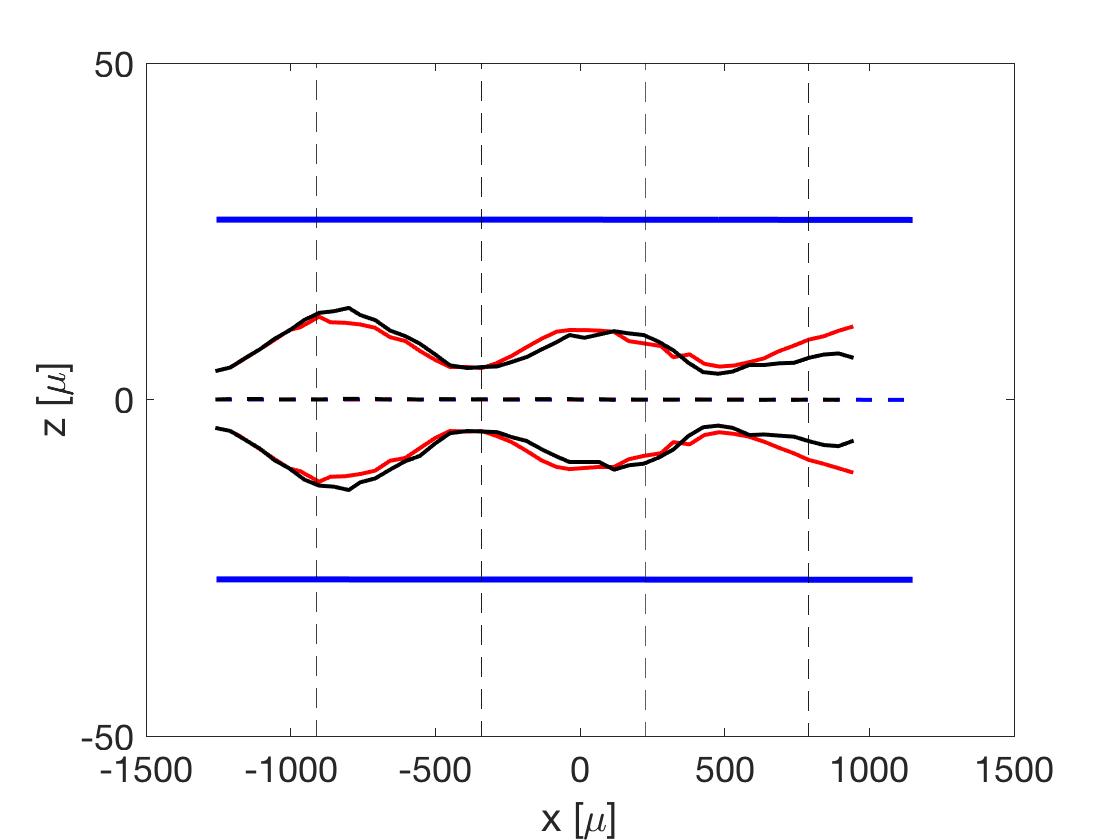}
	\endminipage\hfill
	\minipage{0.32\textwidth}
	\includegraphics[width=\linewidth]{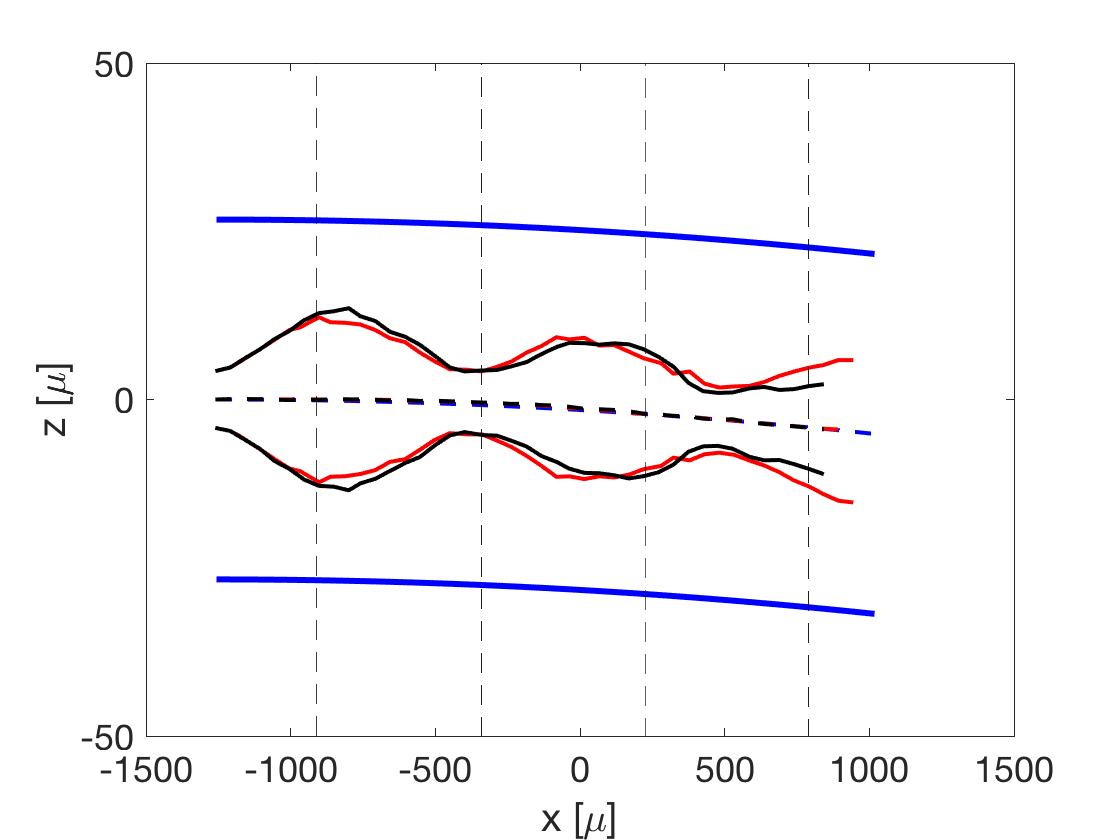}
	\endminipage\hfill
	\minipage{0.32\textwidth}%
	\includegraphics[width=\linewidth]{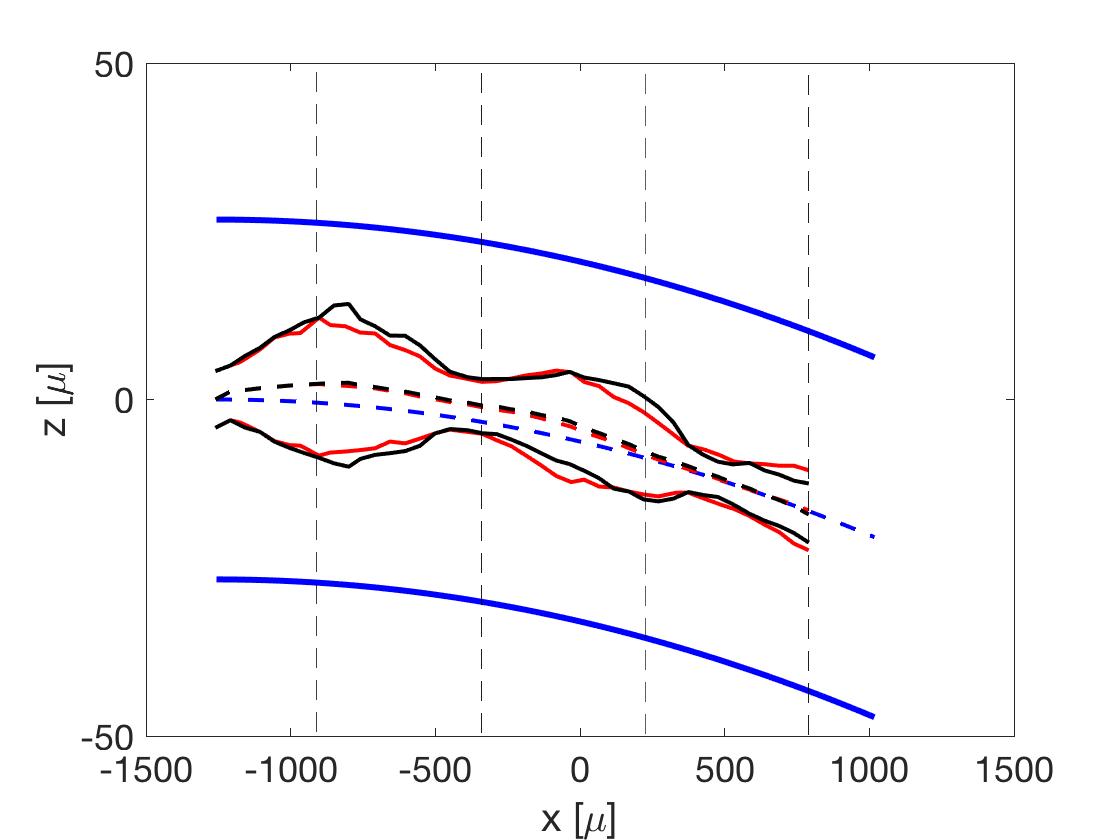}
	\endminipage
	\caption{The envelope of the elaser pulse as it propagates along the plasma channel of a straight capillary (left), 50cm curvature radius (middle) and 12.5cm curvature radius(right). Two laser intensities are shown: $I_0 \approx 10^{18} W/cm^2$ (red)and $I_0 \approx 10^{19} W/cm^2$ (black). The guiding of the high intensity pulse in the 12.5cm curvature radius is evident. }
	\label{threecaps}
\end{figure*}
\begin{figure*}
	\minipage{0.48\textwidth}%
	\includegraphics[width=\linewidth]{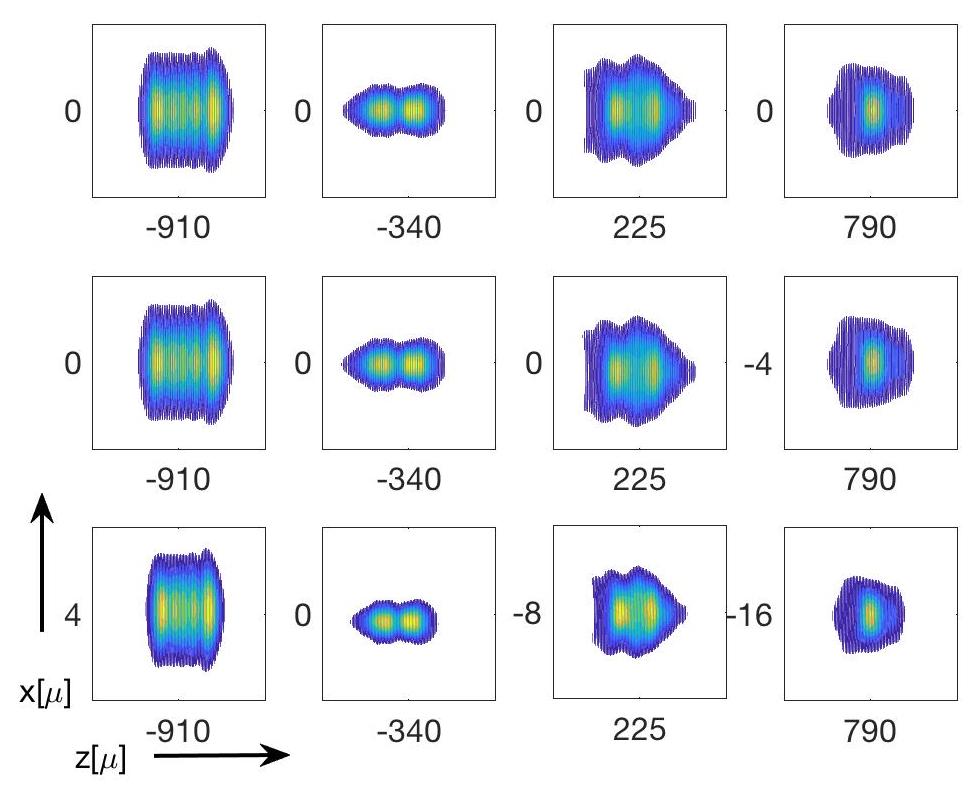}
	\label{fig:awesome_image1}
	\endminipage\hfill
	\minipage{0.48\textwidth}
	\includegraphics[width=\linewidth]{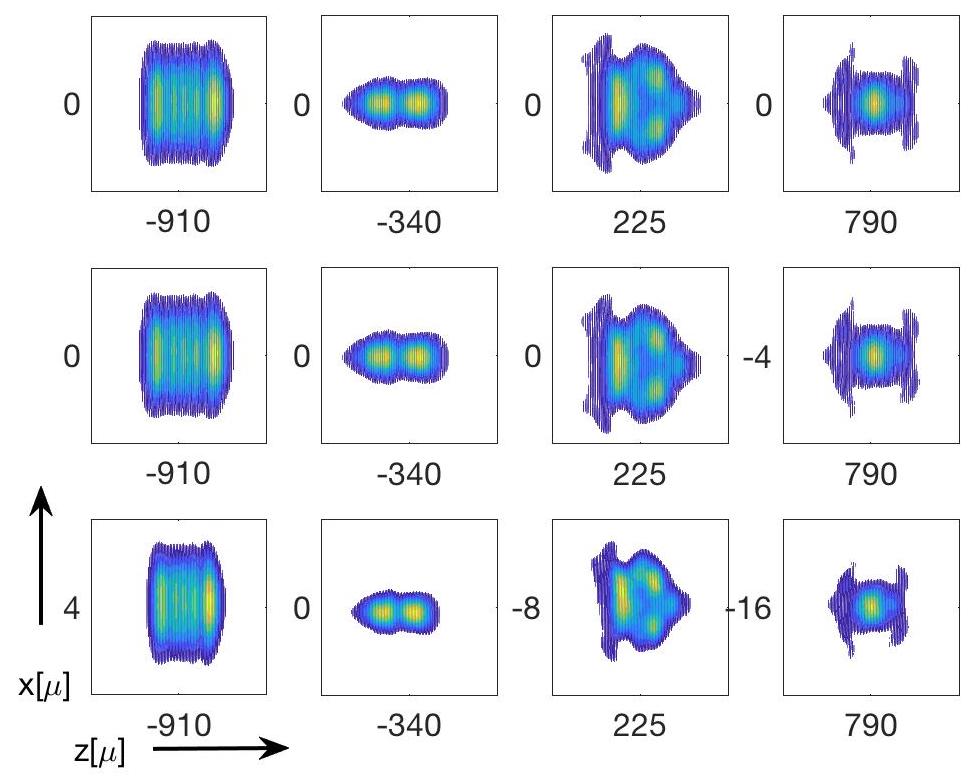}
	\label{fig:awesome_image2}
	\endminipage
	\caption{The laser pulse in four positions along the capillary for two laser intensities. $I_0 \approx 10^{18} W/cm^2$ (left)and $I_0 \approx 10^{19} W/cm^2$ (right). Top row is for straight capillary, middle row for a 50cm curvature radius and botton row for 12.5cm curvature radius. Each box is $20\mu\times20\mu$ with center coordinates given. The four axial positions are described by broken line in Fig. \ref{threecaps}. The energy distribution in the pulse varies as it propagates but it remains confined and guided.}
\end{figure*}
\twocolumngrid
The envelope of a propagating high intensity short pulse lase in a plasma channel is shown in Fig. \ref{threecaps} for three values of curvature radius and two laser intensities.
 The left box in Fig. \ref{threecaps} is for a straight capillary, the middle for a curvature radius of 50cm and the right is for a 12.5cm curvature bent capillary. The axial length of the simulated capillary is 2.5mm. In all cases  red line is for $\approx 10^{18} W/cm^2$ and black line for $\approx 10^{19} W/cm^2$. The guiding of the pulse along the plasma channel is evident even for the 12.5cm curvature radius. Following the propagation of the laser pulse along the plasma channel we found that the tpulse changes its volume in both axial as well as transverse dimentions.
 Fig. \ref{TransverseProfile} demonstrate this for two laser energies $\approx 10^{18} W/cm^2$ on the left and $\approx 10^{19} W/cm^2$ on the right. The three rows are for a straight capillary (top) 50cm curcature (middle) and 12.5cm curvature (bottom). The pulse is plotted in four axial positions (marked in broken lines in Fig. \ref{GuidingRes}). Each figure represents a box of $20\mu\times20\mu$ and the values of the center coordinates are specified. The pulse exhibits a complex behavior along its propagation with transversally and longitudinally change of shape, but it remains confined by the capillary  density gradient. 

In conclusion, we have proposed and experimentally demonstrated a method of concatenating subsequent acceleration stages in a LWFA accelerator. By creating an effective plasma waveguide within curved capillaries, we can couple the new, energetic, replenishing pulse with the main electron acceleration line. In our experiment, this was realized by guiding 800 nm pulses in two curved capillaries; simultaneously at low power and in one capillary at high power (0.2 TW). The guiding was exhibited by comparing the signal of the pulse when traveling through the preformed plasma with the signal measured when no plasma had been created. At low power, the guiding amplified the signal by as much as 800\%, and at high power almost 600\% was reached. Our numerical studies demonstrated that a high intensity short pulse laser can propagate in a plasma channel of small radius of curvature. This significant reduction of the curvature radius is expected to enable shorter capillary junctions. 

\begin{acknowledgments}
This project was partially supported by the BSF and ISF research programs.
\end{acknowledgments}

\bibliography{BibliographyCoupCurvedCap}

\end{document}